\documentclass[fleqn,usenatbib]{mnras}  

\usepackage{newtxtext,newtxmath}

\usepackage[T1]{fontenc}

\DeclareRobustCommand{\VAN}[3]{#2}
\let\VANthebibliography\thebibliography
\def\thebibliography{\DeclareRobustCommand{\VAN}[3]{##3}\VANthebibliography}

\usepackage{graphicx}	
\usepackage{amsmath}	
\usepackage{pdflscape}
\usepackage{bm}

\newcommand{\bea}{\begin{eqnarray} }
\newcommand{\eea}{\end{eqnarray}}

\newcommand{\cloudy}{C{\scriptsize LOUDY} }

\title[Origin of Changing-look AGNs]{Multi-phase gas nature in the sub-pc region of the active galactic nuclei II:  Possible origins of the changing-state AGNs}

\author[K. Wada et al.]{
Keiichi Wada,$^{1,2,3}$\thanks{E-mail: wada@astrophysics.jp}
Yuki Kudoh,$^{4}$
and Tohru Nagao$^{2}$
\\
$^{1}$Kagoshima University, Graduate School of Science and Engineering, Kagoshima 890-0065, Japan\\
$^{2}$Ehime University, Research Center for Space and Cosmic Evolution, Matsuyama 790-8577, Japan\\
$^{3}$Hokkaido University, Faculty of Science, Sapporo 060-0810, Japan\\
$^{4}$Tohoku University, Faculty of Science, Sendai Japan
}

\date{Accepted XXX. Received YYY; in original form ZZZ}

\pubyear{2023}

\begin{document}
\label{firstpage}
\pagerange{\pageref{firstpage}--\pageref{lastpage}}
\maketitle

\begin{abstract}
Multi-wavelength observations of active galactic nuclei (AGNs) often reveal various time scales of variability. Among these phenomena, "changing-look AGNs" are extreme cases where broad emission lines become faint/bright or even disappear/emerge between multi-epoch observations, providing crucial information about AGN internal structures. 
We here focus on "changing-state" AGNs, specifically investigating the transition of optical spectra over years to tens of years.
Based on the axisymmetric radiation-hydrodynamical simulations (Paper I) for the gas dynamics within the dust-sublimation radius, we investigate the spectral properties of ionized gas exposed to the radiation from an AGN with a $10^7 M_\odot$ supermassive black hole.
 We find significant time-dependent variations in the Balmer emission lines by utilizing post-process pseudo-three-dimensional calculations and the spectral synthesis code \cloudy 
 The equivalent width of H$\alpha$ and H$\beta$ changes by a factor of 3,  or the emission lines even disappear during 30 years for the same viewing angle.
The time-dependent behaviour arises primarily from gas dynamics, particularly the formation of non-steady, radiation-driven outflows within the innermost region of the disc ($r  \lesssim10^{-3}$ pc). 
{The intricate interplay between non-spherical radiation sources at the core of AGNs and the dynamic behavior of gas within the dust sublimation radius gives rise to radiation-driven outflows. This non-steady outflow potentially contributes to the observed variability in Balmer line emissions over multi-year timescales in certain AGNs.}

\end{abstract}




\section{Introduction}
Active galactic nuclei (AGNs) often show time variability of their flux across a wide wavelength with a time scale ranging from hours to years \citep[e.g.,][]{ulrich1997, vanden_berk2004}. The variability should reflect the change of the structures/state of the interstellar medium around AGNs and/or the change of the central source itself, i.e., the accretion disc around the supermassive black holes. 
Multi-epoch observations revealed that some AGNs changed their spectral types (i.e., type-1 and type-2). This drastic change is a perplexing phenomenon if the spectral types of AGNs are caused only by obscuration of a parsec-scale ``torus'' as suggested by the unified model \citep{antonucci1985} \citep[see also,]    []{netzer2015, hoenig2019, stalevski2019}. This drastic spectral change is often called "changing-look AGNs" (CL-AGN).

Specifically, the phenomenon is divided into two categories: changing-obscuration AGNs and changing-state AGNs (hereafter CS-AGN)\citep[e.g.,][]{ricci2022}. The former is often observed in X-ray spectra and characterized by variation in $\sim$ hours to days. This change's origin is still unclear, but the short time scale variability implies that the absorber should be compact and dense. On the contrary, CS-AGNs were discovered in repeated observations with an interval of months to tens of years in optical and ultra-violet observations, e.g., NGC 7603 \citep{tohline1976} (see also \citet{ricci2022} and references therein), {and
Mrk 590 \citep{raimundo2019, lawther2022}.}
{
\citet{Wang2023-km} identified nine AGNs with spectral type transitions within $\sim $ 10 yrs out of 59 candidates that show variability in the mid-infra red band. }
Change in the major emission lines is also the case for luminous quasars\citep[e.g.,][]{lamassa2015, runnoe2016}.  Sloan Digital Sky Survey (SDSS) suggested that a few per cent of quasars show disappearing or emerging the broad emission lines in 8-10 years \citep{macleod2016}.
{
The Catalina Real-time Transient Survey identified 111 quasars with specific patterns of optical and mid-infrared spectroscopic change\citep{graham2020}.}

Among nearby changing-look AGNs, NGC 3516 is one of the first AGNs showing strong variability of H$\beta$ lines \citep{andrillat1968} between observations in 1943 and those in 1967. More recently, \citet{shapovalova2019, popovic2023} confirmed that the line profiles of H$\alpha$ and H$\beta$ show a year-scale variability between 1996 and 2021. 
The observed timescales of CS-AGNs, which are typically 1-10 years, may be too short to be explained by a change in the physical state in the accretion disc \citep{ricci2022}. Therefore, the nature of the CS-AGNs would be caused by temporal change of structures in the emission line regions.
{Year-scale variabilities in the column density were also 
suggested from the X-ray spectra in some Seyfert galaxies \citep{2002ApJ...571..234R} (See also \citet{laha2020})}.
{\citet{popovic2023} pointed out that the H$\beta$ line profiles  can be roughly explained with a two-component model, including a disc-like region and a wind originating from the center. This kind of disk-wind structure was, in fact, found in our pc-scale radiation-driven fountain model \citep{wada2012radiation} and also in the central region of the Circinus galaxy \citep{izumi2023}.
\citet{kudoh2023} (Paper I) demonstrated a significant change in the outflow structures around a rotating disk, which motivates us to investigate their spectral properties. }

Recently, the spatial structures of broad line regions (BLRs) were partially resolved using a near-infrared interferometer in some nearby AGNs
\citep{2018Natur.563..657G, gravity2020, 2021A&A...648A.117G}.
The results are consistent with the size determined by the  reverberation mapping (RM) technique
 \citep[e.g.,][]{blandford1982, peterson1993, peterson2004ApJ...613..682P, lawther2018, Baskin2018-hj}.
The outer edge of the BLR is  $\sim 1/3 $ of the dust sublimation radius \citep{netzer_laor1993, suganuma2006, netzer2015, minezaki2019, Netzer2020-hv, GRAVITY_Collaboration2022-md}.  
Suppose the variability is $\sim$ 10 yr, this corresponds to 
the rotational time scale at $r \sim 0.005$ pc around $M_{\rm BH} = 10^7 M_\odot$,
and this is comparable to the outer edge of BLRs is for $L_{AGN} = 10^{43}$ erg s$^{-1}$\citep{netzer2015}. 

 
The hydrodynamics of dusty gas under central radiation has recently been extensively studied, focusing on the "obscuring torus" at a scale of 1--10 pc \citep{wada2012radiation, dorodnitsyn2012, wada2015obscuring, namekata2016, williamson2020}. Multi-dimensional radiation-hydrodynamic simulations have revealed the natural formation of outflowing multi-phase gas with dust, providing a natural explanation for the type-1 and -2 dichotomies in the spectral energy distribution (SED) \citep{schartmann2014}.
This dynamical model, known as the "radiation-driven fountain," effectively accounts for various multi-wavelength observations of the nearby type-2 Seyfert galaxy, the Circinus galaxy, in several aspects. It explains molecular and atomic emission as well as absorption lines in the central 10 pc \citep{izumi2018, wada_fukushige2018, uzuo2021, matsumoto2022, izumi2023}, the conical shape and line ratio properties of the narrow emission line region (e.g., [O III]~$\lambda$5007) \citep{wada_yonekura2018}, and the X-ray spectral energy distribution and lines \citep{buchner2021, ogawa2022}. Notably, the radiation-driven fountain exhibits a time-dependent nature, with the multi-phase gas showing non-steady behavior, particularly for diffuse, ionized gas \citep{schartmann2014}.

The time scale of these changes tends to be shorter in the more central region, even if the global structure, such as the torus-like dusty gas, remains roughly unchanged. This time-dependence adds an additional layer of complexity to the dynamics of the system and has implications for the interpretation of observational data.

In Paper I \citep{kudoh2023}, we found that the dusty and dust-free gas in the central sub-pc region 
is highly time-dependent (see Fig. 8 in Paper I), owing to the non-linear interaction between the non-spherical central radiation field and
the non-uniform gas around the dust-sublimation radius.  The dust sublimation radius itself change on several years. Interestingly, intermittent formation of the shell-like outflows
is formed even for {\it a constant source luminosity}.  

As the second paper of the series, we here focus on optical lines and their time-evolution derived from the gas inside the dust-sublimation radius using a high-spatial-resolution radiation-driven fountain model.

Following \citet{wada_yonekura2018}, we analyze snapshot sof the hydrodynamic simulation using 
the photo-ionized code C{\scriptsize LOUDY} \citep{ferland2017}. 
We here focus on evolution of line profiles of the Balmer lines.
This is an attempt to understand the origins of the year-scale variability of emission lines based on a physics-motivated 
multi-dimensional model in the central region of AGNs.

\section{Numerical Methods}

\subsection{Physical model in gas, dust, and radiation} \label{sec:model}
We use several snapshot data from a axisymmetric radiation-hydrodynamic simulation in a quasi-steady state (see Paper I in detail) and
calculate the radiative-transfer as a post process (see \S 2.2). Here, we briefly summarize the hydrodynamic model.

\begin{figure}
\centering
\includegraphics[width = 8cm]{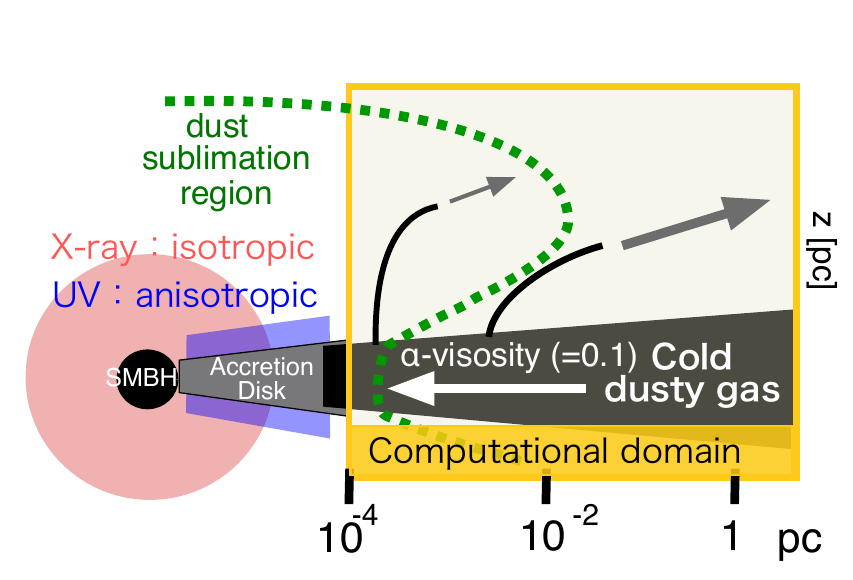}

\caption{Model setup:  The central radiation field is
assumed to be anisotropic and isotropic for
UV and X-ray, respectively. Therefore, 
the dust sublimation region is not spherical (see Paper I in detail). The dusty, cold gas is
supplied through the disc and the outflow is launched from the inner most region of the disc ($r \lesssim 0.1$ pc). 
}

\label{wada_fig: model}
\end{figure}

We solve the evolution of a dusty gas disc with mass inflow irradiated by a central source in a computational box of $r = 10^{-4} \sim 50$ pc
(Figure \ref{wada_fig: model}). 
This is an extension of the three-dimensional radiation-driven fountain simulations \citep{wada2012radiation,wada2015obscuring} with a higher resolution.  However, we assume an axisymmetric distribution using a cylindrical coordinate.
The included physics here are those of radiative heating by X-ray as well as the radiation force that induces both dusty and ionized gases.
The black hole mass is $M_{\rm BH} = 10^7 M_\odot$ and the Eddington ratio is 0.1 (the bolometric luminosity is $1.25 \times 10^{44}$ erg s$^{-1}$).

The basic equations are
\begin{eqnarray}
\frac{\partial \rho}{\partial t} + \bm{\nabla} \cdot \left[\rho \bm{v} \right]  =0,  
\label{eq:mass}
\end{eqnarray}
\begin{eqnarray}
 \displaystyle \frac{\partial \rho \bm{v}}{\partial t}
 + \bm{\nabla} \cdot \left[ \rho  \mathbf{vv} + {P}_{\rm g} \mathbf{I}  \right]  
 =  \bm{f}_{\rm rad} + \bm{f}_{\rm grav} + \bm{f}_{\rm vis} ,
\label{eq:momentum}
\end{eqnarray}
\begin{eqnarray}
\displaystyle \frac{\partial e}{\partial t}
+ \bm{\nabla} \cdot \left[ \left( e+ P_{\rm g}  \right) \bm{v}  \right]
 = - \rho {\cal L} + \bm{ v} \cdot \bm{f}_{\rm rad} +  \\ \nonumber 
  \bm{ v} \cdot \bm{f}_{\rm grav} + W_{\rm vis},
 \label{eq:energy}
\end{eqnarray}
where total energy density is $e=P_{\rm g}/(\gamma-1)+\rho v^2/2$, and the specific heat ratio obtained adiabatically, i.e.  $\gamma =5/3$. 
${\cal L}$ is the net heating/cooling rate per unit mass.
We adopted the gravitational force, $\bm{f}_{\rm grav} \equiv  -\rho G M_{\rm BH} \bm{e}_{r} /r^2 $, where $G$ denotes the gravitational constant and $r=\sqrt{R^2+z^2}$ is the distance from the center of BH.
The radiation force is $\bm{f}_{\text{rad}} \equiv \int \nabla \cdot F_{\nu} \bm{e}_{r} d \nu$, where $F_{\nu}$ is the radiation flux.

We assume  the $\alpha$ viscosity  to achieve mass accretion through the disc 
as $\nu_{\rm vis}= \alpha c_s^2/\Omega_{\rm K}$ where the viscosity depends on the sound speed $c_s$ and the Keplerian angular speed $\Omega_{\rm K}$ \citep{shakura_sunyaev1973}.
The viscous force in Equation (\ref{eq:momentum}) and the viscous heating in Equation \ref{eq:energy} are taken from \citet{osuga2005}:

\begin{equation}
  \bm{f}_{\rm vis} \equiv \frac{\bm{e}_{\varphi}}{R^2} \frac{\partial}{\partial R} \left[ R^2 \alpha P_{\rm g} \frac{R^2}{v_{\varphi}} \frac{\partial}{ \partial R } \left( \frac{v_{\varphi}  }{R} \right)  \right], 
\end{equation}
and
\begin{equation}
  W_{\rm vis} \equiv \alpha P_{\rm g} \frac{R}{v_{\varphi} } \left[  R \frac{\partial}{ \partial  R } \left( \frac{v_{\varphi}}{R} \right)  \right]^2.
\end{equation}
We assumed the viscosity parameter $\alpha$ to obtain the gas supply around the disc mid-plane,
\begin{equation}
\alpha=
\left\{\begin{array}{ll}
0.1 \quad &  n > 10^{3} \text{ cm}^{-3} ~\&~ T_{\textrm{g}}< 10^3 \text{ K} \\ 
0.0 \quad & {\rm otherwise} 
\end{array}\right.
\end{equation}

We consider heating by UV and X-ray \citep{maloney1996, meijerink2005, wada2012radiation} and the optically thin radiative cooling \citep{meijerink_spaans2005, wada_papadopoulos2009}.
We assume that dust grains sublimate above a dust sublimation temperature $T_{\text{sub}}=1500$ K.
Dust temperature is generally used in local thermal equilibrium with radiation.

We use the public MHD code CANS+ \citep{matsumoto2019}
\footnote{\url{https://github.com/chiba-aplab/cansplus}} 
with the additional module to evaluate the radiation force and radiative heating/cooling. However, we ignore the magnetic field in the present model.
The number of computational cells in each direction is set to $(N_R, N_z) = (1200, 2304)$.
The cell sizes in the uniform region give high resolutions $\Delta R=\Delta z=5 \times 10^{-5}$ pc for $R< 3.5 \times 10^{-2}$ pc and $|z|< 3.26 \times 10^{-2}$ pc.
On the outside, the cells are stretched to approximately 0.1 pc for a maximum simulation box of $R=|z|=50$ pc (Figure \ref{wada_fig: model}).

\begin{figure*}
\centering
%
\includegraphics[width =17.0cm]{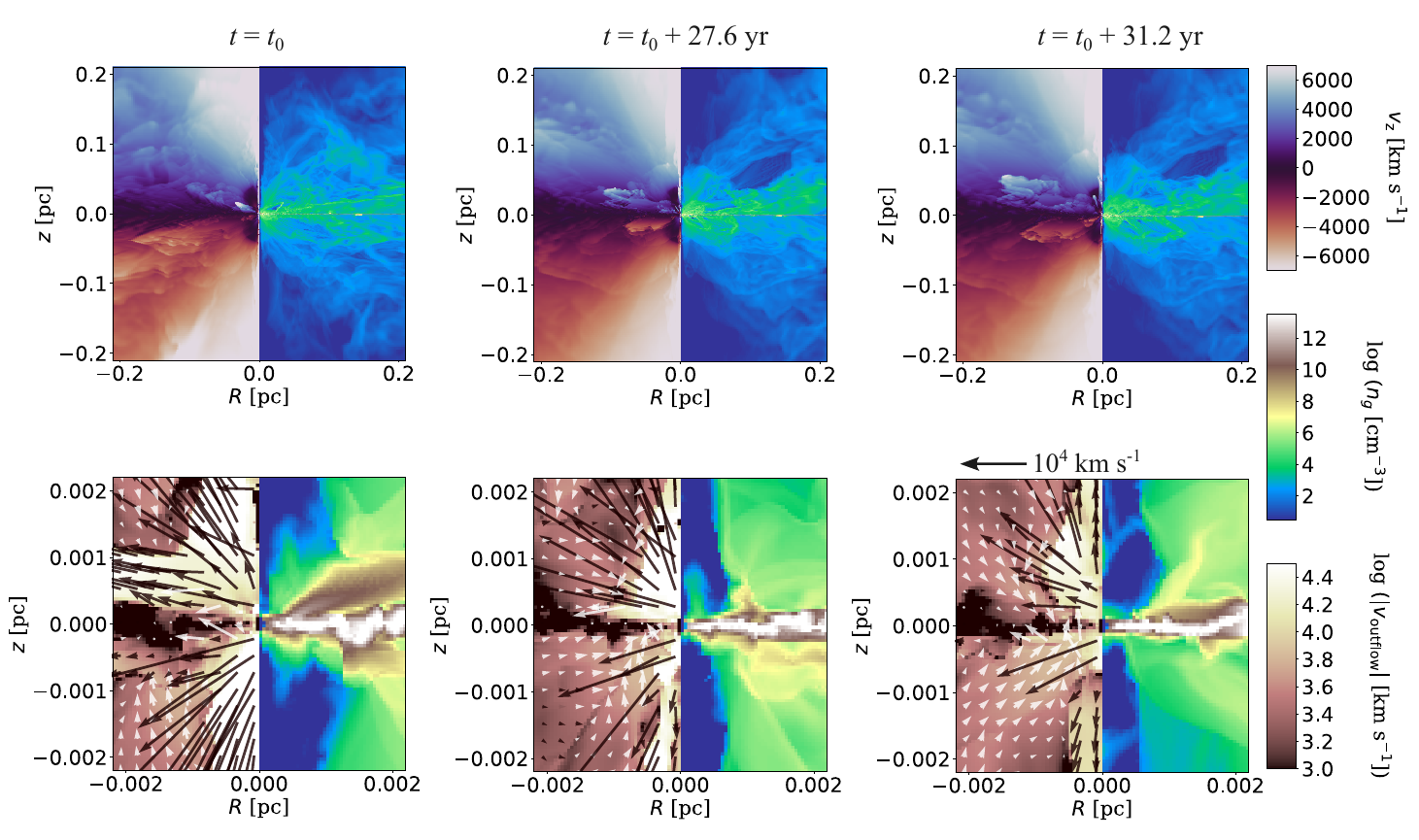} 

\caption{Three snapshots of the radiation-hydrodynamic simulations,
Three snapshots are shown ($t = t_0,  t_0+27.6\, {\rm yr}, t_0+31.2 \, {\rm yr}$)
from which the spectra (Fig. 3) are calculated.
In each pane,  the top and bottom are the same, but for the central 0.2 pc and 0.002 pc regions, respectively. 
See also Paper I.
In the top three figures, the left-half panel shows $z$-component of the gas velocity $v_z$. The right-half panel shows the number density of the gas.}

\label{wada_fig: hydro}
\end{figure*}

\subsection{Radiative transfer using C{\scriptsize LOUDY}}

For the spectrum calculations, we used density, temperature, and velocities in the central $r \le 0.02$ pc in the hydrodynamic simulation described in Section 2.1
(Figure \ref{wada_fig: hydro}), and the data of the 2D cylindrical coordinate
were modified for polar grid cells with $ (N_r, N_\theta) = (400, 41)$ for $ -30 \le \theta \le 30^\circ$, where $\theta$ is the angle from
the equatorial plane.
We then ran the spectral synthesis code C{\scriptsize LOUDY} (version 17.03) \citep{ferland2017}.

The SED of the central source was derived from 
 \textsf{Cloudy}'s \textsf{AGN} command and is represented as
\begin{eqnarray}
\label{eqn:agncon}
F_\nu    =  \nu ^{\alpha _\mathrm{UV} } \exp \left( { - h\nu /kT_\mathrm{BB} } \right)\exp \left( { - kT_\mathrm{IR} /h\nu } \right)\cos{i} \nonumber\\
                         +  a\nu ^{\alpha_\mathrm{X} } \exp \left( { - h\nu /E_1 } \right) \exp \left( { - E_2 /h\nu } \right),
\end{eqnarray}
where 
$\alpha _\mathrm{UV} = -0.5$, $T_\mathrm{BB} = 10^5$~K, 
$\alpha_\mathrm{X} = -0.7$, $a$ is a constant that yields the X-ray-to-UV ratio $\alpha_\mathrm{OX} = -1.4$, 
 $kT_\mathrm{IR} = 0.01$~Ryd,
$E_\mathrm{1} = 300$~keV, $E_\mathrm{2} = 0.1$~Ryd, and $i$ is the angle from the z axis (i.e., rotational axis).
The UV radiation (first term), which derives from the geometrically
thin optically thick disc, 
was assumed to be proportional to $\cos{i}$. 
By contrast, the X-ray component (second term) was assumed to be isotropic (Figure \ref{wada_fig: model}).

We assume that grains are sublimated (i.e., no grains) in the data used in C{\scriptsize LOUDY} ,
and the Solar metallicity is assumed. 
 The following are parts of the input file for C{\scriptsize LOUDY}:
\begin{verbatim}
abundances "default.abn" no grains
grains ism function sublimation
filling factor 1.0
no molecules
set nend 2000
set continuum resolution 0.2
\end{verbatim}
{The wavelength resolution corresponds to $\sim$ 200 km s$^{-1}$ for 6000 \AA.}

The transmitted SED, calculated using \cloudy for the inner most cell,
was used as an incident SED for the next outward radial cell,
and this procedure was repeated up to the outer edge (i.e., $r=0.02$ pc) for a given radial ray \citep[see][in detail]{wada_yonekura2018}.

 Upon completion of all \cloudy calculations, we  {\it observed} the system (i.e., all the grid cells within 
 $r = 0.02 $ pc) along the 
 line of sight, assuming the viewing angle ($\theta_v = 0$ means
face-on).

 {
Although the basic equations (Section  \ref{sec:model}) are solved in the $R-z$ plane, we assume that the system is axisymmetric in the azimuthal direction (often referred to as the '2.5-dimensional' treatment) to calculate spectra.
By assuming angular momentum conservation, we determine the azimuthal velocity ($v_\phi$) in the gravitational potential.
We consider three-dimensional velocities in each cell (i.e., $v_R, v_z,$ and $v_\phi$) to calculate Doppler-shifted emission lines using the Cloudy spectrum for a given viewing angle.
We assume 256 azimuthal cells covering $2\pi,$ and the spectrum from all the grid cells is integrated.
Note that under this assumption, there are no non-axisymmetric structures in the gas. Therefore, each emission line is expected to be roughly symmetric with respect to the systemic velocity.
}

%
\section{Time evolution of the optical spectra}
%

Figure \ref{wada_fig: spect1} shows evolution of the obtained spectra between 4000\AA  ~and 7000\AA ~in 31.2 yrs, assuming a viewing angle  $\theta_v = 5^\circ$ from the rotational axis.  Here we used the central 100 radial grid cells ($r < 0.005$ pc) to generate the spectra\footnote{We confirmed that including the data at $r > 0.005$ pc do not affect the optical emission lines discussed here.}.  
The spectra are generated for the upper half (i.e., $z > 0$) of the computational box, assuming that the dense gas disc near $z = 0$ (see Fig. 2) is optically thick.  For $\theta_v = 175^\circ$, see Fig. \ref{wada_fig: spect2} below. 
All the Balmer lines shown here are the strongest at $t = t_0$, when a prominent outflow is generated in the central region for $z > 0$. The outflow seen in Fig. \ref{wada_fig: hydro}a is transient, and getting less prominent after 30 yrs. 
In the two spectra separated with $\sim 4$ yrs, H$\beta$, H$\delta$ and H$\gamma$ are weaker than those at $t= t_0$. 
Note that the continuum levels are also slightly different, but this is not caused by the change of the 
central source, which is assumed constant during the calculation, but owing to the propagation of the radiation through the non-uniform gas.


\begin{table}
\caption{{Equivalent widths and the line ratios 
for $\theta_v = 5^\circ $(left numbers) and $\theta_v = 175^\circ$ (right numbers) in the three snapshots. }}
\label{tab:example}
\begin{tabular}{lccc}
\hline
$t - t_0$ & ${\rm H}\alpha (5^\circ | \, 175^\circ)$ &  ${\rm H}\beta (5^\circ | \, 175^\circ)$  &  ${\rm H}\alpha / {\rm H}\beta (5^\circ | \, 175^\circ)$ \\
 yr &\AA  & \AA  &   \\
\hline
$0$ & 31\, | 8.0 &  10.0 | 2.1 & 1.5  | 2.0 \\
$27.6$ & 3.9 | 5.4 &  \, 0.8 | 0.4 &  1.3 | 1.5 \\
$31.2$ & 8.9 | 3.5 &    1.0 |  --- &  4.5 | ---  \\
\hline
\end{tabular}
\end{table}

{Figure \ref{wada_fig: spect2} shows detail profiles and time difference of 
H$\alpha$ and H$\beta$.  In each snapshot, two spectra  for  $\theta_v = 5^\circ$ or $175^\circ$ are shown for the data above and below the disc plane (i.e., $z > 0$ and $z < 0$, respectively). 
In both lines-of-sight, the full width at half maximum (FWHM) is largest for the snapshot associated with strong outflowing gas at $t = t_0$;
FWHM $\sim 4000$ km s$^{-1}$.   As seen in Fig. \ref{wada_fig: hydro},  the structures of the gas in the central region is not symmetric against $z = 0$, and they are time-dependent.
FWHM at $t = t_0$ for $z < 0$  (the dotted line) is $\lesssim 2000$ km s$^{-1}$, which is in contrast to the spectrum for $z > 0$.
The narrow component originates in rotating gas of the upper part of the disc, where the gas density is $10^8-10^{12}$ cm$^{-3}$. 
{In Table 1, the equivalent widths (EW) of H$\alpha$ and H$\beta$, 
and the line ratio are summarized for the three snapshots.
At $t = t_0$,  Balmer lines are the strongest among the three snapshots.
After $27.2$ yr, EWs of both lines becomes $\sim 1/10$ .
Four years later, the lines becomes stronger again.
However, the line width remains narrow (FWHM of H$\alpha$ is $\sim 1000$ km s$^{-1}$). }

\begin{figure}
\centering
\includegraphics[width = 9.0cm]{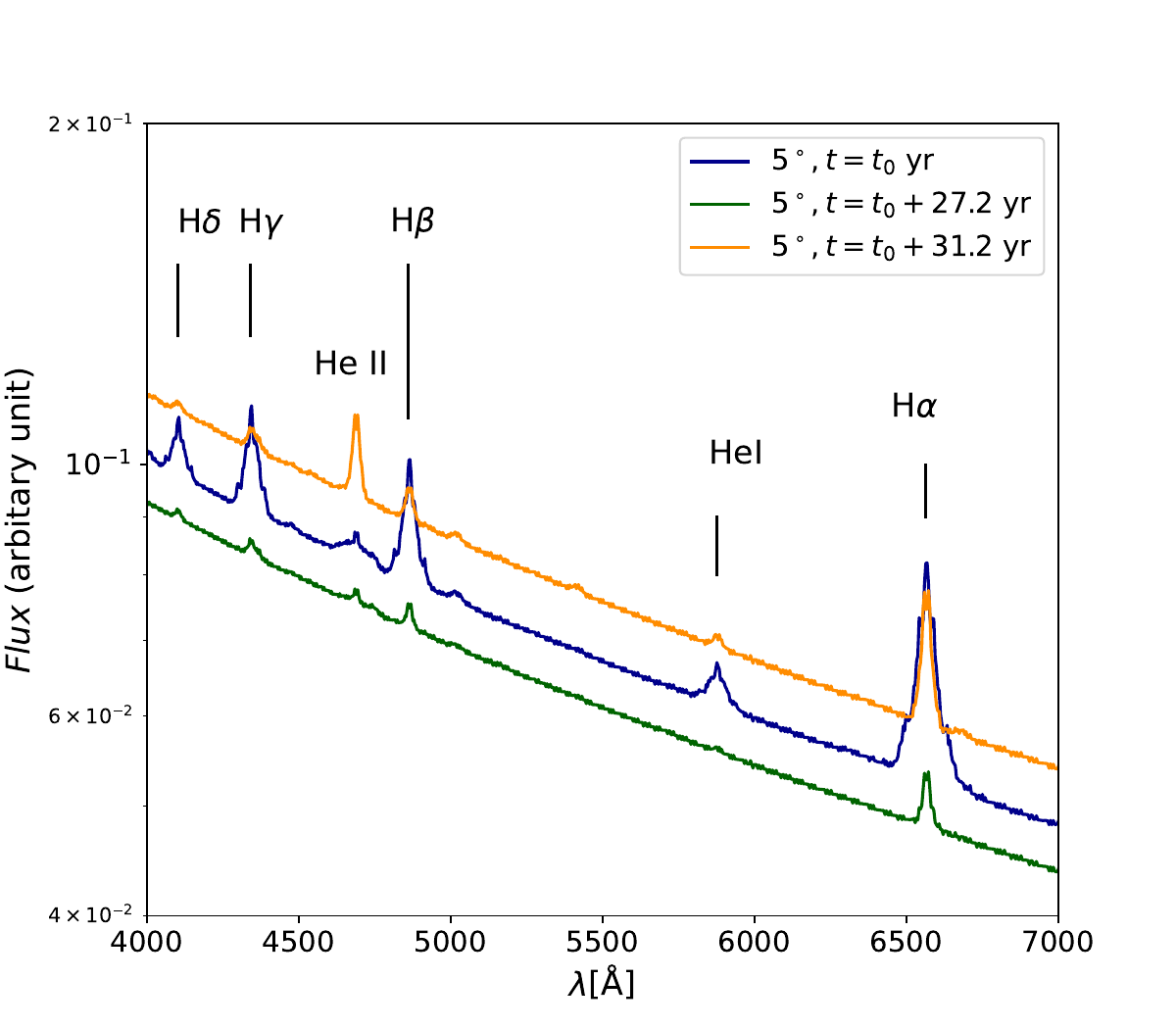} 
\caption{Time evolution of spectra  of H$\alpha$, and H$\beta$.
Three snapshots are shown ($t = t_0,  t_0+27.6\, {\rm yr}, t_0+31.2 \, {\rm yr}$), where $t_0$ is the
snapshot showing the strong outflow (Fig. 2a).
The viewing angle is  $5^\circ$ (i.e., nearly face-on) and, only for the upper ($z > 0$, solid lines) half of the computational box 
is used to calculate the spectra.
}
\label{wada_fig: spect1}
\end{figure}

\begin{figure}
\centering

\includegraphics[width = 8cm]{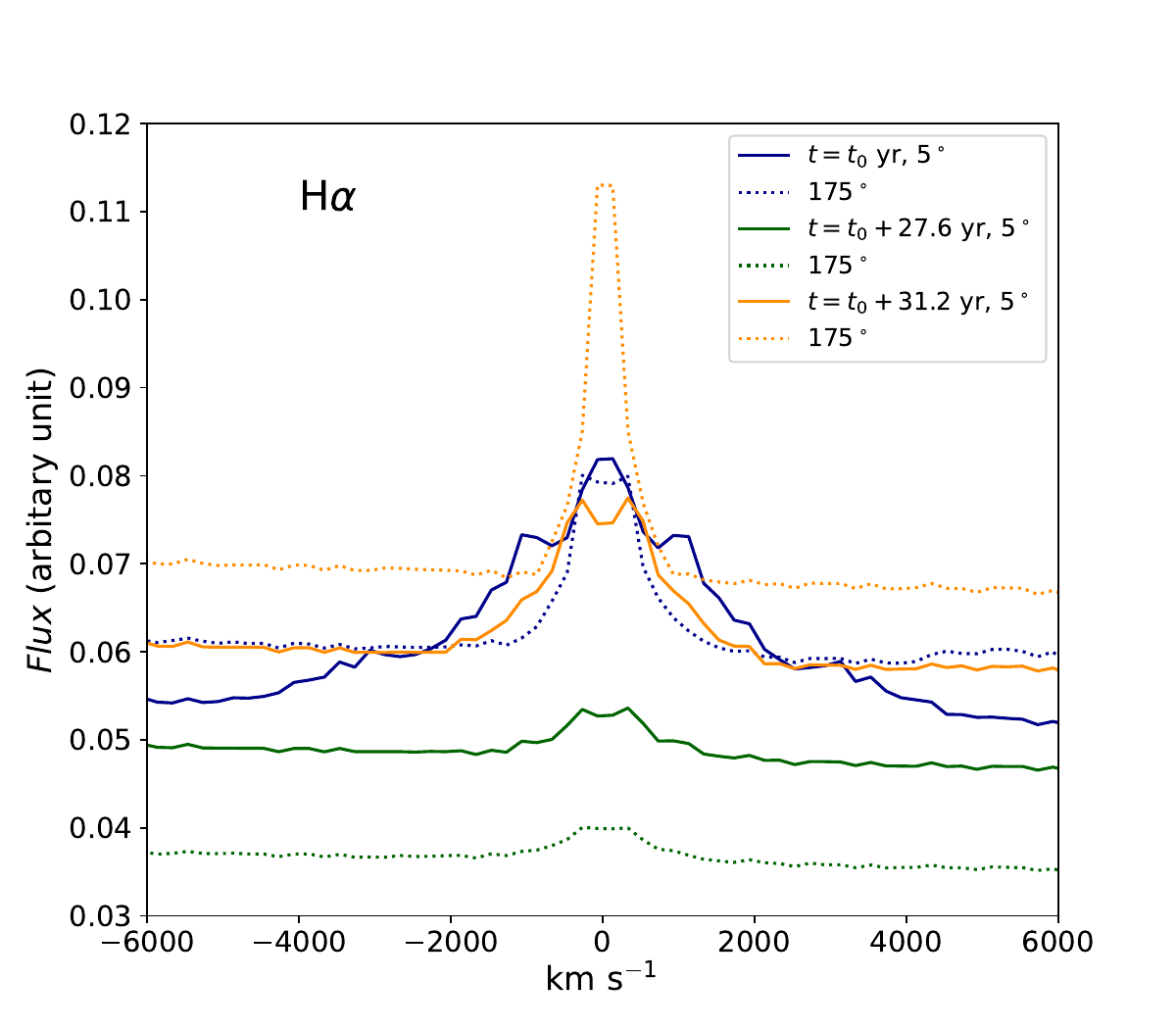}  
\includegraphics[width = 8cm]{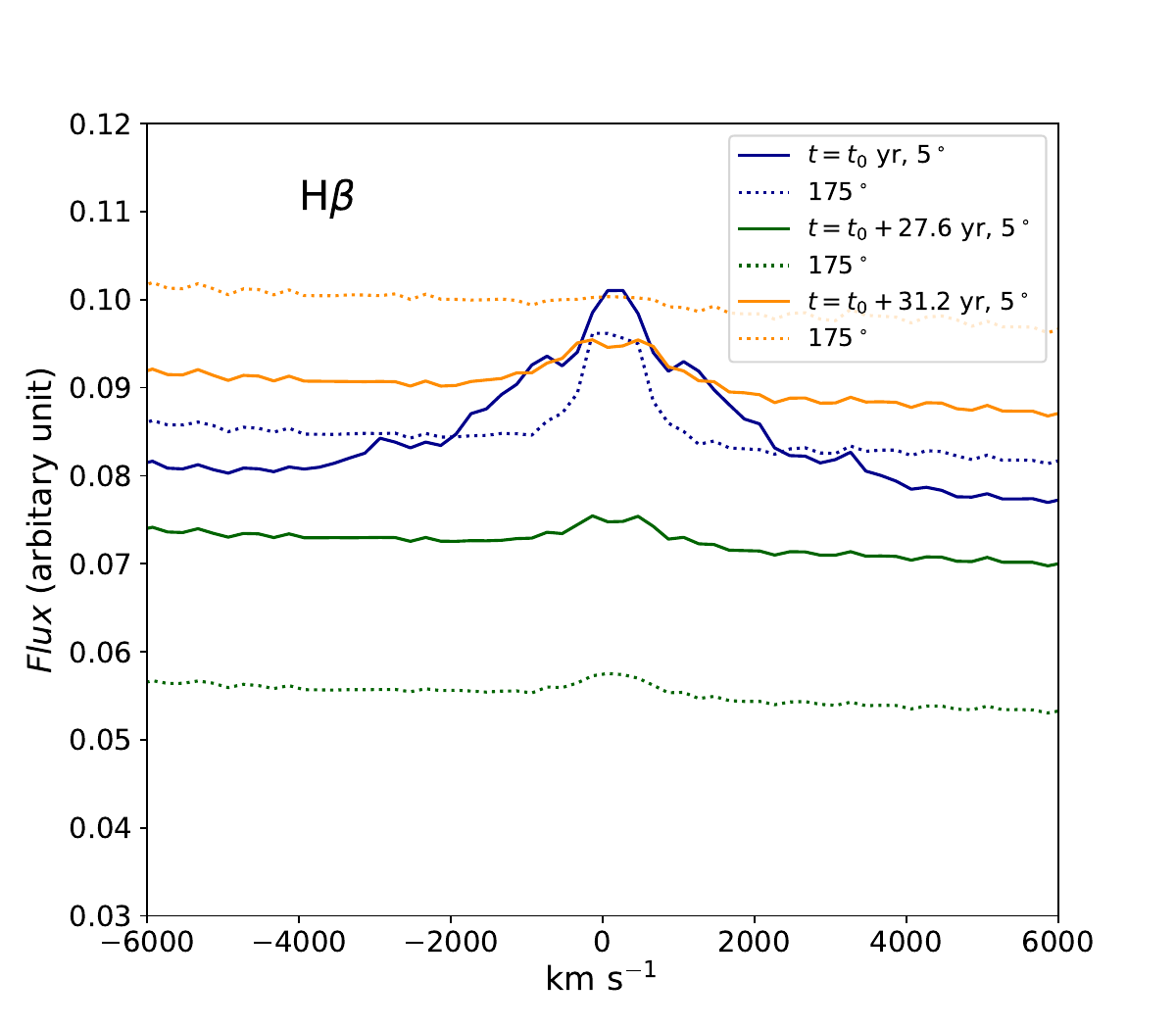}  

\caption{{Same as Fig. 3, but for  H$\alpha$ and H$\beta$.
In each snapshot, two spectra are obtained for the upper ($z > 0$, solid lines) and lower half ($z < 0$, dotted lines) of the computational box, assuming the viewing angles, $5^\circ$ (i.e., nearly face-on).
 Note that the fluctuating profiles are partially
 owing to the coarse velocity resolution ($\sim $200 km s$^{-1}$).  
}}

\label{wada_fig: spect2}
\end{figure}

\begin{figure}
\centering
\includegraphics[width = 8cm]{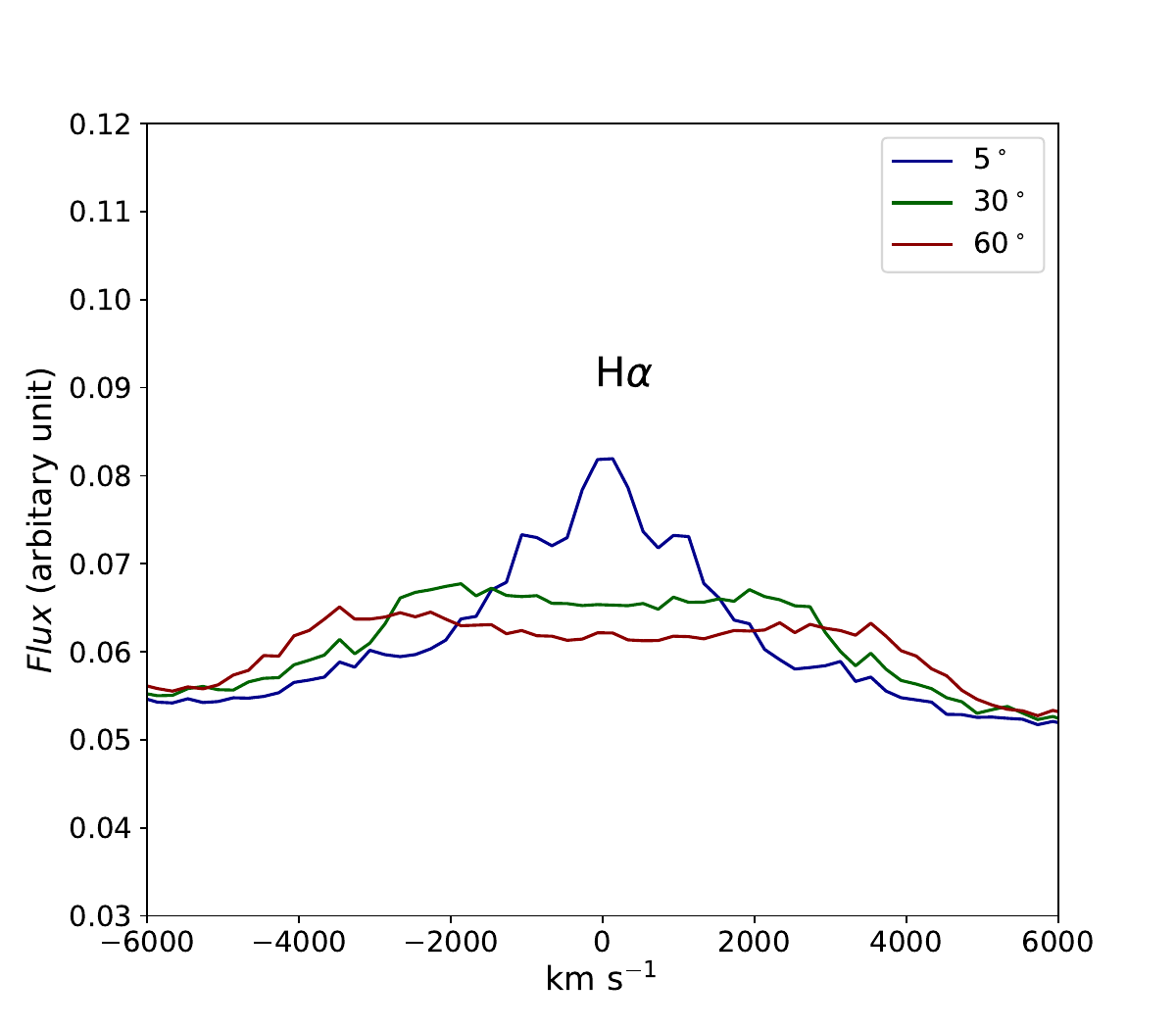} 

\caption{H$\alpha$ line profiles at $t = t_0$ are plotted for $\theta_v = 5, 30,$ and 60$^\circ$ from the rotational axis. {These profiles are the upper half of the computational box (i.e. for $z > 0$).}}

\label{wada_fig: spect4}
\end{figure}

Figure \ref{wada_fig: spect4} compares H$\alpha$ profiles at $t = t_0$ for three viewing angles: $\theta_v = 5, 30$, and  $60^\circ$. 
It shows that the profiles are wider for larger viewing angles.  The two peaks in the profile for $\theta_v =  60^\circ$ correspond to
the fast outflowing gas. 
Note that we assumed in this paper that the velocity and density structures seen in Fig.  \ref{wada_fig: hydro} are the same for all the azimuthal angle. 
Therefore, 
one does not expect to observe a velocity shift
against the line center (see also section 4.3).

Suppose the line width reflects gravity in the central region,  we expect that
FWHM $\sim \sqrt{M_{\rm BH} G/R} \sim 4700$ km s$^{-1} (M_{\rm BH}/10^7 M_\odot)^{1/2} (R/0.002 \, {\rm pc})^{-1/2}$. 
{However, Fig. \ref{wada_fig: spect4}  shows that FWHM for $\theta_v = 60^\circ$ is $\Delta V \sim 8000$ km s$^{-1}$, which is much larger than the value typically observed in type-1 Seyfert galaxies. 
However, the nuclear emission for this large viewing angle may be obscured by a parsec-scale material (e.g., the torus), which is not considered in the present analysis.}

%
\section{Discussion} 
%
\subsection{Origin of the variability of the spectra}
The observed changes in optical spectra in Section 3 result from the dynamic interaction between the non-steady gas within the central region and the AGN's radiation field. In our previous work (Paper I), we extensively discussed the dynamics of the dusty and dust-free gas within the dust sublimation radius. Most outflows within this region are not fast enough to escape the central sub-parsec region and eventually fall back to the disc, leading to gas infall towards the central region through the disc. Consequently, the sublimation radius experiences fluctuations on timescales of a few years, with these changes being most pronounced near the disc surface (see Fig. 8 of Paper I).

This circulation of gas closely resembles the concept of a "radiation-driven fountain," which was proposed as a mechanism to create a geometrically thick torus-like structure on a parsec scale \citep{wada2012radiation, wada2015obscuring}. In this process, both X-ray heating and radiation pressure play vital roles in uplifting the gas from a thin disc. However, within the central $10^{-3}$ pc, the radial motion of the gas dominates, and we observe intermittent formation of dense outflows, as depicted in Fig. \ref{wada_fig: hydro}a. This dynamic behavior highlights the complex and ever-changing nature of the gas within the central region, influenced by the intricate interplay between radiation forces and the gas dynamics.
}

{The dust sublimation region is not spherical where the dust temperature exceeds the limit ($\sim 1500$ K) (see the green dotted line in Figure \ref{wada_fig: model}). 
This is because of the non-spherical UV radiation field and the attenuation caused by the dense gas in the disc.
The dust-free gas is launched near the dust sublimation limit, where the gas pressure and radiation force  contribute to the acceleration of
outflows. The disc gas close to the SMBH becomes geometrically thin due to stronger gravity \citep[e.g.,][]{wada_norman2002}. On the other hand, the acceleration by the 
radiation pressure becomes weaker beyond the dust sublimation limit because of the attenuation of the disc gas itself. 
As a result, the disc gas near the dust sublimation limit can be most effectively outflowing.}

{
The intricate multiphase structure and time variability observed in our simulations naturally arise from the complex interplay between the nonspherical central radiation field and the nonuniform distribution of dusty and dust-free gases. The outward propagating shocks play a crucial role in peeling off the cold, dense gas from the disc surface, leading to a phase transition from cold, neutral gas in the disc to hot, ionized gas through shock-heating. This process, in addition to the direct heating and ionization by the central radiation, contributes significantly to the overall dynamics of the system.

Importantly, even if the central radiation remains constant as assumed in this paper, the attenuation of radiation by the gas within the central region of approximately $10^{-3}$ pc is time-dependent, reflecting structural changes in the disc and the generation of outflows. These outflows are not steady but are repeatedly formed, creating multi-shell structures reminiscent of a "lotus-flower" pattern within the parsec-scale region (see Fig. 2 and 3 in Paper I).

While the outflows originate in the innermost region of the disc ($< 10^{-3}$ pc), once the gas in this region is expelled, it takes roughly a dynamical time for the radius to replenish due to mass accretion through the disc plane. As the outflows propagate outward, they become less dense than the typical broad-line region (BLR) gas. Moreover, the central radiation is attenuated both by the outflowing gas itself and by the disc gas. As a result, the broad emission line component eventually disappears within several years. This process highlights the intricate interplay between the central radiation, the gas dynamics, and the formation of outflows, contributing to the observed spectral variability in AGNs.
}

\subsection{Comparison with observations}
One of the well observed nearby CS-AGNs in a long period is NGC 3516, 
which shows strong variation in intensity of the broad lines\citep[][references therein]{shapovalova2019}.
Although the black hole  mass of this galaxy is $4.73\pm 1.4 \times 10^7 M_\odot$ \citep{shapovalova2019} is factor of five 
larger than the present model, it is worth to compare the observations with the results. 

\citet{shapovalova2019}  monitored this galaxy from 1996 to 2018, and 
reported years or sub-year change of H$\alpha$ and H$\beta$ lines. 
NGC 3516 in 2007 showed broad H$\alpha$ and H$\beta$ spectra, but
it shows that H$\beta$ almost disappeared in 2014, and H$\alpha$ became five-times weaker than that in 2007.
\citet{popovic2023}  analyzed change of H$\beta$ line profile of the same galaxy in details from 1996 to 2021.
They showed that H$\beta$ became stronger again from 2019-2021.  
They concluded that {\it the change from type-1 to type-2 of NGC 3516 is most likely caused by intrinsic effects in the BLR.}
This year-scale change of  spectra is 
consistent to what we found in this paper, and indeed the line strength and shapes change
depending on the physical state and dynamics of the gas inside the dust-sublimation radius.

 \citet{popovic2023}  reported that the line profiles are mainly consisted of two components: line cores and the broad line wings.
They called the former component "intermediate line region" (ILR) and the latter "disc-like BLR".  
They also noticed that the ILR component is shift to the blue and it is more shifted in the low activity phase than 
in the high activity phase. This is not actually observed in our results;
The line centers were not shifted.
In the three snapshots in 31.2 yrs, the broadest line width is observed when the EWs of the Balmer lines are largest 
at $t = t_0$, i.e., the outflow dominated for $z > 0$, 
and the narrow components always presented.  We rather interpret that the narrow components are originated in 
the relatively stable "disc".

One may noticed that the H$\alpha$ and H$\beta$ line profiles in NGC 3516 and also in our model are {\it not smooth}.
This is in contrast to the smooth Balmer line profiles in some AGNs \citep{arav1998, laor2006}.
Our results suggested that some sub-structures of the line profile, such as the "shoulders" seen in the H$\alpha$ profile Fig. \ref{wada_fig: spect4} , are in fact originated in the outflows. Note that the velocity resolution here is 200 km s$^{-1}$.

However, it  would be premature to make detail comparison in terms of the sub-components of line profile  and
their time-evolution between observations and models, because our axisymmetric model should be much 
simpler than the real structures, which should be three-dimensional (see Fig. \ref{wada_fig: spect4} and discussion below).


\subsection{Toward more realistic model of the changing-state AGNs}
{
In Figure \ref{wada_fig: spect1}, we observe variations in spectral structures for the gas above and below the equatorial plane in our axisymmetric model. This finding suggests that the line profiles could also undergo changes in the azimuthal direction. Recent studies have proposed the presence of sub-structures within BLRs \citep{wang2022}, implying that line profiles might exhibit additional complexity due to rotational effects. \citet{wang2022} explored line profiles based on a rotating disc model with spiral features, which could potentially be detected through high-quality reverberation mapping and interferometric observations like those conducted by the Very Large Telescope Interferometer (VLTI). Understanding the connection between the structure and dynamics of the dust sublimation region and the observed variability in emission lines could shed light on the nature of the BLRs.

To study substructures of the emission lines, comprehensive three-dimensional radiation-hydrodynamics and radiative transfer calculations are essential for the gas within the region between the dust sublimation radius and the accretion disc. 
Previous studies utilizing three-dimensional "radiation-driven fountain" models on a pc -10 pc-scale have demonstrated the existence of non-axisymmetric structures in the outflow \citep{wada2012radiation, wada2015obscuring}.
Thus full three-dimensional modeling for the nuclear sub-pc region is crucial to unravel the complexities associated with the broad-line variations. It is important to note that while the outflows may account for the broad lines, they cannot solely explain any observed blue- or redshifts in the line profiles.

In further advancing our understanding, we should explore a broad parameter space within the physics-based numerical models, including variations in $M_{\rm BH}$, the Eddington ratio and SEDs. Such investigations will offer valuable insights into the intricate dynamics of AGNs and contribute to comprehending the origin and characteristics of their emission lines.
}

\section{Conclusion} 
Multi-wavelength observations of active galactic nuclei (AGNs) often reveal various time scales of variability. Among these phenomena, "changing-look AGNs" stand out as extreme cases where broad emission lines drastically change between muti-epoch observations, providing crucial information about AGN internal structures. In this study, we focus on "changing-state" AGNs, specifically investigating the transition of spectra in the optical band over years to tens of years.

Building upon previous work (Paper I), which employed axisymmetric radiation-hydrodynamical simulations to explore the gas dynamics within the dust-sublimation radius, we investigate the spectral properties of ionized gas exposed to the radiation from an AGN featuring a supermassive black hole with a mass of $10^7 M_\odot$. Utilizing post-process pseudo-three-dimensional calculations and the spectral synthesis code C{\scriptsize LOUDY} \citep{ferland2017}, we find significant time-dependent variations in the Balmer emission lines. Specifically, the equivalent width of H$\alpha$ and H$\beta$ changes by a factor of 3 during a 30-year period, and the emission lines even disappear in some snapshots for the same viewing angle.

{
The time-dependent behavior primarily arises from gas dynamics, especially the formation of non-steady, radiation-driven outflows in the dust sublimation region (Paper I).
The complex interaction between non-spherical radiation sources in the center of the AGN and gas dynamics within the dust sublimation radius forms radiation-driven outflows. This non-steady outflow may partly account for the year-scale Balmer line variability observed in some AGNs.
Importantly, this time variation, attributed to the non-steady outflows, can occur even when the luminosity of the central source remains constant over the period of the variation. In other words, the CS-AGN phenomenon might be caused not only by changes in the accretion state but also by dynamic phenomena outside the accretion disk.
Our findings shed light on the underlying mechanisms driving the variability of changing-state AGNs and underscore the importance of radiation feedback in shaping the spectral characteristics of these enigmatic objects.}

As \citet{Baldwin1995} indicated in their ``locally optimal cloud'' representation, the emission line spectrum can be reproduced by 
integrating various properties of emitting clouds, and the spectra do not necessarily represent 
physical conditions such as pressure, gas density, or ionization of individual clouds.
They also speculated that a chaotic cloud environment could be the source of the lines, and therefore the lines 
reflect mostly global properties of the clouds.  {In our first attempt to reproduce the observed broadband spectral properties, we find that this somewhat disordered representation is supported by a radiation-driven fountain model. This suggests that a 3-D hydrodynamic model with sufficient spatial resolution is needed for a comprehensive understanding of the origin of BLR.}

\section*{Data Availability}
The data underlying this article will be shared on reasonable request to the corresponding author.
\section*{Acknowledgements}
We appreciate many constructive comments and suggestions from the referee.
We thank G. Ferland and the \cloudy team for their regular support. 
Numerical computations of the radiation-hydrodynamic model were performed on a Cray XC50 at the Center for Computational Astrophysics at the National Astronomical Observatory of Japan and using the Fugaku supercomputer at RIKEN. This work was supported by JSPS KAKENHI Grant Number 21H04496 (KW) and 23H01215 (TN).
The work used computational resources of Fugaku provided by RIKEN through the HPCI System Research Project (Project IDs: hp210147, hp210219).



\bibliographystyle{mnras}
\bibliography{Paperpile-AGN-Sep02,agn_papers,agn_papers_takasao}


\bsp	
\label{lastpage}
\end{document}